**Title: Quantification of free-living activity patterns using accelerometry in adults with mental illness**

Running title: Activity patterns in adults with mental illness

Authors: Justin J. Chapman (1), James A. Roberts (1,2), Vinh T. Nguyen (1), Michael Breakspear (1,3)

((1) Systems Neuroscience Group, QIMR Berghofer Medical Research Institute, (2) Centre for Integrative Brain Function, QIMR Berghofer Medical Research Institute, (3) The Royal Brisbane and Women's Hospital)

**Correspondence to:**

*Name*:    Justin Chapman

*Post:*    QIMR Berghofer Medical Research Institute

           Herston, Queensland, Australia 4029

*Email*:   justin.chapman@qimrberghofer.edu.au

*Phone:*   +61 7 3362 3369

**24 pages;** *Main text*: 4,486 words; *Abstract*: 200 words

Five (5) *tables* are referred to in the text, two of which are supplementary.

Seven (7) *figures* are referred to in the text, one of which is supplementary.

Physical activity is disrupted in many psychiatric disorders. Advances in everyday technologies – such as accelerometers in smart phones – opens exciting possibilities for non-intrusive acquisition of activity data. Successful exploitation of this opportunity requires the validation of analytical methods that can capture the full movement spectrum. The study aim was to demonstrate an analytical approach to characterise accelerometer-derived activity patterns. Here, we use statistical methods to characterize accelerometer-derived activity patterns from a heterogeneous sample of 99 community-based adults with mental illnesses. Diagnoses were screened using the Mini International Neuropsychiatric Interview, and participants wore accelerometers for one week. We studied the relative ability of simple (exponential), complex (heavy-tailed), and composite models to explain patterns of activity and inactivity. Activity during wakefulness was a composite of brief random (exponential) movements and complex (heavy-tailed) processes, whereas movement during sleep lacked the heavy-tailed component. In contrast, inactivity followed a heavy-tailed process, lacking the random component. Activity patterns differed in nature between those with a diagnosis of bipolar disorder and a primary psychotic disorder. These results show the potential of complex models to quantify the rich nature of human movement captured by accelerometry during wake and sleep, and the interaction with diagnosis and health.

# INTRODUCTION

Mobile smart phones are ubiquitous in modern society, and their numerous sensors, such as GPS (Global Positioning System), gyroscopes and accelerometers, offer a unique opportunity to gather rich empirical data on free-living activity patterns[1-5]. Smart phones have been used to facilitate monitoring of early warning signs of relapse in patients with bipolar disorder (e.g. self-reporting sleep patterns)[6], and smart phone accelerometers can be used to discern mood changes by monitoring the frequency and duration of bursts of activity[7]. Mobile technologies are crucial in the development of 'telehealth' systems, enabling remote interaction between patients and clinicians, and automated monitoring of symptoms[8-11]. Accelerometry also has potential in monitoring physical health, by measuring total activity and the distribution of sedentary bouts, which have been shown to be major risk factors for adverse physical health outcomes[12]. The opportunity to obtain rich behavioral data, using non-invasive methods, has elevated smart phone employment to a priority position in several large-scale international research efforts[13-17]. There is a pressing need to explore more sophisticated analytical techniques for quantifying mental illness symptoms using accelerometry[18].

Previous use of accelerometry in psychiatric research has established proof-of-principle utility, but has predominantly only analyzed basic properties of the available data. For example, accelerometry has been used to distinguish subtypes of schizophrenia using simple summary statistics, namely the mean of activity counts[7,19-22]. Patients with schizophrenia appear to exhibit more structured activity than healthy controls or people with depression, as quantified by inter-daily stability and intra-daily variance[23], but increased disorder at shorter time scales[24]. Increased dynamical entropy at short timescales has also been found in people with schizophrenia and bipolar disorder[25]. Over longer times, bipolar disorder is associated with higher intra-daily variability and lower inter-daily stability[26]. Bipolar patients hospitalized during either manic or depressive episodes have different activity patterns quantified using measures of variability[27], and circadian rhythms are related to mood variation in outpatients with euthymic bipolar disorder[28]. A systematic review reported that patients with depression have a lower daytime activity and higher night-time activity than individuals without depression, and concluded that analytical methods need to be improved to extract all relevant features from actigraph data[18]. Whilst informative, use of summary statistics fails to capture the rich complexity of human behavioral data stretching across many orders of magnitude, from a brief motion (e.g. a foot tap), to lengthy outcome-oriented endeavors (e.g. purposeful walking). Capturing the breadth of active and inactive behaviors requires techniques that quantify the entire spectrum of movement[29].

Human activity can be conceptualized as a combination of internally-driven and external cue-triggered actions. Internally-driven actions are voluntary, purposeful actions, whereas externally-triggered actions are executed in response to dynamic environmental stimuli[30]. Although these components – volitional versus reactive – appear to arise from distinct neural processes occurring in different regions of the brain[31], most free-living activities are motivated by a combination of internal and external influences. Given the broad spectrum of influences that could potentially determine the timing of any action, one might assume that free-living activity can be characterized by random, uncorrelated statistical processes (i.e., Poisson statistics). Recent work in quantifying the waiting times of inter-human communications, however, points toward non-Poisson processes governing email communication[32], web browsing[33] and printing requests[34]. Such activities are characterized by bursts of rapidly occurring events interspersed with longer waiting times, resulting in a so-called heavy-tailed distribution of the waiting time between tasks[35]. This suggests that other types of human activity, such as physical movement, might be of a non-Poisson nature.

Recent work building on this approach has revealed novel differences in movement patterns of adults with mental illness. In a small but intriguing study, Nakamura et al. demonstrated that the distributions of active or inactive periods in healthy adults followed distinct classes of non-Poisson models[36]. They showed that active distributions followed a stretched exponential (Weibull) form, implying that we become "trapped" in active states: that is, the longer we are active, the less likely we are to cease movement at any particular moment. In contrast, inactive distributions followed a power-law form, implying that inactive periods do not have a characteristic time scale. Moreover, the inactive distributions in patients with depression followed the same power-law form as healthy adults, but with an altered scaling parameter, indicating more frequent episodes of longer resting periods[29,36]. Further, people with schizophrenia exhibited an enhanced persistency for *both* inactivity and activity when compared with controls[37]. This method has also been used to identify motor retardation in depressed patients, with a higher scaling parameter describing patients with motor retardation[38].

Accelerometry thus has substantial potential to detect clinically relevant changes in movement patterns. However, accelerometer-based studies to date have been limited by two important factors: Small sample size and the absence of robust statistical methods to compare movement patterns between mental illnesses. Moreover, analyses have employed only simple models and hence neglected the possibility that everyday activity reflects a composite of different modes of activity. Thus, the systematic validation of non-Poisson statistics in patterns of activity measured using accelerometry has not been achieved, and requires application to a large cohort with a spectrum of mental illnesses. This is a crucial prelude to broader translation of movement data acquired using mobile phones and other smart devices. Here, we demonstrate the utility of a rigorous quantitative analysis incorporating novel stochastic models for characterizing activity patterns in adults with mental illness.

# METHODS AND MATERIALS

## Participants

One hundred and fourteen participants were recruited from psychiatric outpatient clinics and community-based mental health organizations in Brisbane, Australia. Participants were ambulatory, English speaking, and over 18 years of age, who self-identified as having a mental illness. Informed written consent was obtained from all participants. Ethical approval for the study was obtained from The University of Queensland Behavioral and Social Sciences Human Ethical Review Committee (2012000908), and the Royal Brisbane & Women's Hospital Human Ethical Review Committee (HREC/12/QRBW/286). These methods were carried out in accordance with the approved guidelines.

Of the 114 consenting participants, 13 did not complete the study. Reasons for withdrawal included lack of time (n=2), anxiety/paranoia about the accelerometer (n=3), forgetfulness (n=2), self-perceived inability to adhere to the study (n=2), and hospitalization (n=2); two participants were withdrawn by clinical staff, because they considered their client's mental health not suitable for participation. Data from two participants who completed the protocol were unable to be used because of an accelerometer fault (n=1), or failure to follow the protocol (n=1); data from 99 participants were therefore included in the analyses. The mean age was 40 (SD=11; range 18-74), and 48% were female. Formal assessment revealed considerable diagnostic heterogeneity: 83 participants screened positive for at least one of the 13 major diagnoses on the MINI-Plus, and 42 met criteria for two or more diagnoses. Positively screened diagnoses included anxiety (n=33), psychoses (n=31), bipolar disorder (n=23), substance dependence (n=23), and depression (n=20). The cohort thus represents a typically heterogeneous population of community-dwelling adults with mental illness. Diagnostic information is provided in Supplementary Tables 1 and S1.

**Table 1**
**Participant health and demographic characteristics (n=99)**

| | |
|---|---|
| Age; mean (SD) | 40.5 (11.3) years |
| | range=18-71 years |
| Female; n (%) | 47 (48%) |
| *Psychological distress* [d] | |
| mean (SD) | 14.8 (5.5) |
| High distress; n (%) | 43 (43%) |
| *Number of current diagnoses* | n (%) |
| 1 | 41 (41%) |
| 2 | 33 (33%) |
| 3-5 | 9 (9%) |
| None | 16 (16%) |
| *Single current diagnosis* [a] | n (%) |
| Psychoses | 15 (15%) |
| Bipolar disorder | 14 (14%) |
| Anxiety | 5 (5%) |
| Substance use | 4 (4%) |
| Depression | 3 (3%) |
| None [c] | 16 (16%) |
| *Multiple current diagnoses* [b] | n (%) |
| Anxiety | 33 (33%) |
| Psychoses | 31 (31%) |
| Bipolar disorder | 23 (23%) |
| Substance dependence | 23 (23%) |
| Depression | 20 (20%) |
| *BMI (kg/m$^2$)* [e] | n (%) |
| <18.5 | 2 (2%) |
| 18.5 – 24.9 | 19 (19%) |
| 25 – 29.9 | 31 (31%) |
| >30 | 47 (48%) |

[a] Current psychiatric diagnosis screened using the Mini International Neuropsychiatric Interview (MINI-Plus). Only participants with a single diagnosis are presented, hence, the proportions do not add to 100%. A more detailed list is provided in Supplementary Table S1.

[b] All positively screened diagnoses are presented for participants with multiple psychiatric comorbidities, hence the proportions add to more than 100%.

[c] Self-reported diagnoses of participants who did not screen positive for a current psychiatric diagnosis on the MINI-Plus were: psychoses (n=6), depression (n=4), depression and anxiety (n=2), depression and bipolar disorder (n=1), anxiety (n=1), bipolar disorder (n=1), and psychosis secondary to an acquired brain injury (n=1).

[d] Psychological distress in the previous four weeks measured using the Kessler-6 scale; scores range from 6 to 30, scores over 15 indicate high distress.

[e] Height and weight were measured using a stadiometer and electronic scales. BMI was calculated as [weight (kg)] / [height (m)].

**Data acquisition**

Height and weight of all participants were measured during the first visit, and a Mini International Neuropsychiatric Interview (MINI-Plus) was conducted to screen for 13 diagnoses. The MINI-Plus interview covers major depressive disorder (MDD), panic disorder (PD), agoraphobia, obsessive-compulsive disorder (OCD), post-traumatic stress disorder (PTSD), substance dependence (drug/alcohol), psychotic disorders (schizophrenia, schizoaffective disorder, schizophreniform disorder, psychotic disorder NOS), bipolar disorder 1 and 2 (BP1 and BP2), anorexia nervosa, bulimia nervosa, generalized anxiety disorder (GAD), and adult attention deficit hyperactivity disorder (ADHD). The MINI-Plus interview has been shown to have a sensitivity of 0.45-0.96, and specificity of 0.45-0.98 for these disorders, compared with the Structured Clinical Interview for DSM-IIIR[39]. Participants were asked to wear an ActiGraph GT3X+ accelerometer (ActiGraph, Pensacola, FL) on the right hip 24 hours/day for seven consecutive days. Acceleration data were sampled from the three axes at 30 Hz. Participants were asked to remove the device only to go in water (e.g. shower, swim), and to record the time to bed, time out of bed, and accelerometer non-wear times, in a diary.

**Data Pre-processing**

Raw acceleration data were filtered at a bandwidth of 0.25 to 2.5 Hz corresponding to the timescales of human movement[40]. Data were converted to counts per second (cps), and the vector magnitude was used to estimate the relative intensity of activity for each 1-second epoch. Data were also converted to counts per minute (cpm) for comparison with previous research, and daily averages of the proportion of time spent in sedentary behaviour (SB) and moderate-to-vigorous activity (MVPA) were calculated using validated thresholds: ≤100 counts per minute (cpm) and >2019 cpm, respectively[41]. Accelerometer non-wear time was identified from participant diaries, and from consecutive zero counts for 180 minutes or longer, and removed. Participants' self-reported bed times were used to define their waking and sleeping data. Analyses were performed on pooled waking and sleeping data, and compared with analyses for waking and sleeping data alone. Analyses were conducted in Matlab 2016a (The MathWorks, Inc., Natick, Massachusetts, United States).

Participants wore the monitor for a median of 23.4 hours/day (IQR=16.6 to 23.9). The median proportion of self-reported time spent awake, to time spent asleep, was 1.6 (IQR=1.3 to 1.9). When considering the proportion of time that the monitor was worn in either waking hours or sleeping hours, the median ratio of these two proportions was 1.00 (IQR=0.98 to 1.19), indicating an approximate balance between waking and sleeping data. Participants with insufficient sleep data were only included in the analysis of waking data (defined as a ratio of awake to sleep data greater than 3; ten (10%) cases had insufficient sleep data).

**Data analysis**

We reduced free-living activity data to a sequence of *active events*, and *inactive waiting-times* ($T_{A_i}$ and $T_{W_i}$, respectively). These series represent accelerometer data above, and below, a threshold defined by the mean of non-zero activity counts (Figure 1 and Figure 2)[29]. The respective time series are defined by $T_{A_i} = t_{D_i} - t_{U_i}$, and $T_{W_i} = t_{U_{i+1}} - t_{D_i}$, where $t_{U_i}$ is the beginning (upward threshold crossing), and $t_{D_i}$ is the end (downward threshold crossing) of the $i^{th}$ event.

Cumulative probability distributions (CDFs) were derived from these active and inactive time series. CDFs characterize the entire activity spectrum of a system, and their functional form reflects the nature of the underlying generative process[35]. We first fitted the CDFs with a power-law model using a robust fitting procedure, in which the Kolmogorov-Smirnov statistic is minimized, and the log-likelihood is maximized, over all possible $x_{min}$ values, and a range of likely model parameters ($\alpha$=1.1 to 5.0, in steps of 0.01)[42]. Previous communications have cautioned against using limited sets of competing models for observed data[43] – to compare the relative successes within a more comprehensive set, the optimal $x_{min}$ value for the power-law fit was used to fit four other standard competing models: *i)* truncated power-law; *ii)* simple exponential; *iii)* stretched exponential; and *iv)* log-normal. In addition to these simple and long-tailed models, we also fitted two composite models: *v)* bi-exponential, and *vi)* sum of exponential and truncated power-law models, representing a composite of underlying processes. All models were fitted to the data using the maximum likelihood estimation technique, which maximizes the agreement of the model with the observed data[44]. Table 2 gives an overview of these models, and a description of the processes they represent.

More complex models have more parameters, and thus may 'over fit' the data. The Bayesian information criterion (BIC) is a measure of goodness-of-fit which penalizes each model's relative complexity, thus favoring more parsimonious models. We used the BIC to determine the best fitting model for each subject, because it applies a stronger penalty to model complexity than other model selection statistics, such as the Akaike Information Criterion (AIC). We then used a Bayesian model selection routine to determine the most likely model to describe the entire cohort. Each competing model's exceedance probability (φ) was calculated, representing the probability that the model generated data from a randomly chosen subject within the group[45]. This method of model selection has been shown to be more robust than fixed-effects analysis, or conventional frequentist tests of model evidences, particularly in the presence of outliers[45]. The *group-averaged* parameters of the successful models were then calculated as the average of each subject's model parameters, weighted by the renormalized model evidences of the plausible models[46].

**Table 2**
**Overview of model distributions**

| Model name | PDF function | Description |
|---|---|---|
| Power-law | $Cx^{-\alpha}$ | Power-law distributions arise in empirical data when an observable results from an underlying scale-invariant process, quantified by the scaling exponent ($\alpha$). Example: earthquake magnitudes follow a power-law distribution[65]. |
| Exponential | $Ce^{-\lambda x}$ | Exponential distributions occur widely in nature in the waiting time between events from a random statistical process. Such events are Markovian ('memoryless'), occurring independent of the time since the last event, and have a characteristic time scale characterized by $\lambda$. Example: call center arrivals[66]. |
| Log-normal | $C \dfrac{1}{x} \exp\left[\dfrac{-(\ln x - \mu)^2}{2\sigma^2}\right]$ | A probability distribution with a normally distributed logarithm defined by its mean ($\mu$) and standard deviation ($\sigma$), which arises from the multiplication of many positive, independent random variables (cf. the normal distribution which arises from the *sum* of independent random variables via the central limit theorem). Log-normal distributions describe situations of growth, where the growth rate is independent of size. Examples: axon diameters and firing rates of cortical neurons[67]. |
| Weibull | $Cx^{\beta-1}e^{-\lambda x^\beta}$ | Defined by the *shape* ($\beta$) and the *scale* ($\lambda$), Weibull distributions (also known as *stretched exponential*) are commonly used in survival and reliability analyses. The probability of an observable over time decreases for $\beta<1$, is constant for $\beta=1$, and increases for $\beta>1$. Example: wind-speed distributions[68]. |
| Truncated power-law | $Cx^{-\alpha}e^{-\lambda x}$ | Power-laws truncated at longer time scales by an exponential distribution are indicative of scale-free phenomena limited by the finite physical extent of the system (i.e., events have an upper limit). Example: bursts in neonatal EEG[69]. |
| Biexponential | $C_1\delta e^{-\lambda x} + C_2(1-\delta)e^{-\gamma x}$ | The result of two independent underlying exponential processes, with their own characteristic time scales defined by $\lambda$ and $\gamma$. Example: bimodal alpha power[70]. |
| Sum of exponential and truncated power-law | $C_1\delta e^{-\gamma x} + C_2(1-\delta)(x^{-\alpha}e^{-\lambda x})$ | A composite distribution formed by the sum of an exponentially-truncated power-law and an exponential distribution. |

PDF: Probability Distribution Function
Normalization constants ($C$, $C_1$, $C_2$) are given in Clauset et al.[42], or calculated numerically where a closed form solution does not exist.

Individual participant characteristics may influence which model is most successful for each participant, and the model parameters of the successful group-wise model. We investigated the influence of i) health and demographic characteristics: age, sex, BMI, psychological distress, smoking status (ex-smoker>6 months, occasional, daily); and ii) summary statistics of each participant's accelerometer data: MVPA and SB. To investigate the influence of these characteristics on model selection, we performed a non-parametric Kruskal-Wallis test between participants with different successful explanatory models for active and inactive periods (determined by their BICs). To investigate predictors of the model parameters, hierarchical multiple regressions were performed using participant characteristics to predict the model parameters of the successful group-wise models for active and inactive periods.

**RESULTS**

Acceleration data show a striking diurnal cycle of increased day-time activity and decreased (but not absent) night activity (Figure 1). The corresponding active and inactive time series show characteristic burst-like activity (Figure 2), with the sporadic appearance of high amplitude (lengthy) events, suggesting that standard exponential models are unlikely to compete well against candidate heavy-tailed models for these data.

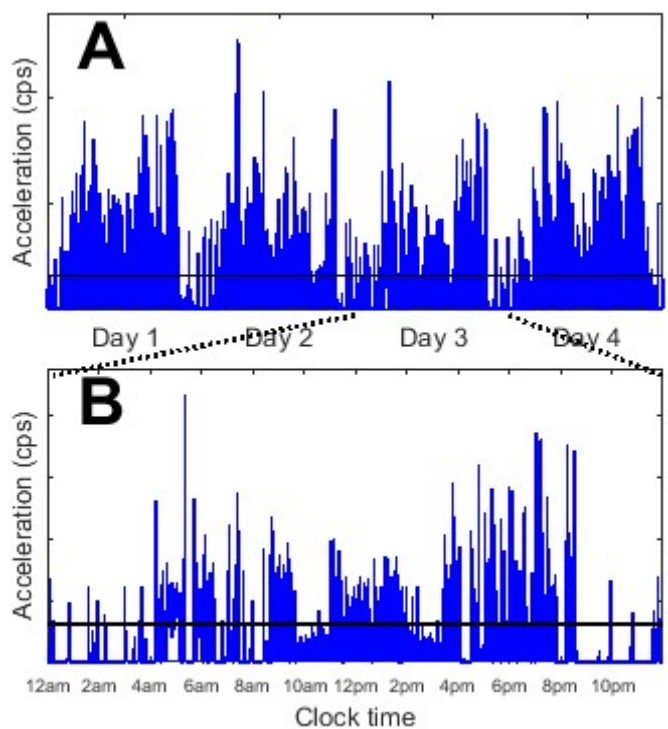

**Figure 1** –Accelerometer data averaged over 1-second epochs (counts per second: cps) from a single subject. (**A**) Four consecutive days. (**B**) Close-up of Day 3 beginning and ending at midnight; time in 24-hour format. The horizontal line represents the threshold (defined as the mean of non-zero activity counts) used to derive the active and inactive time series.

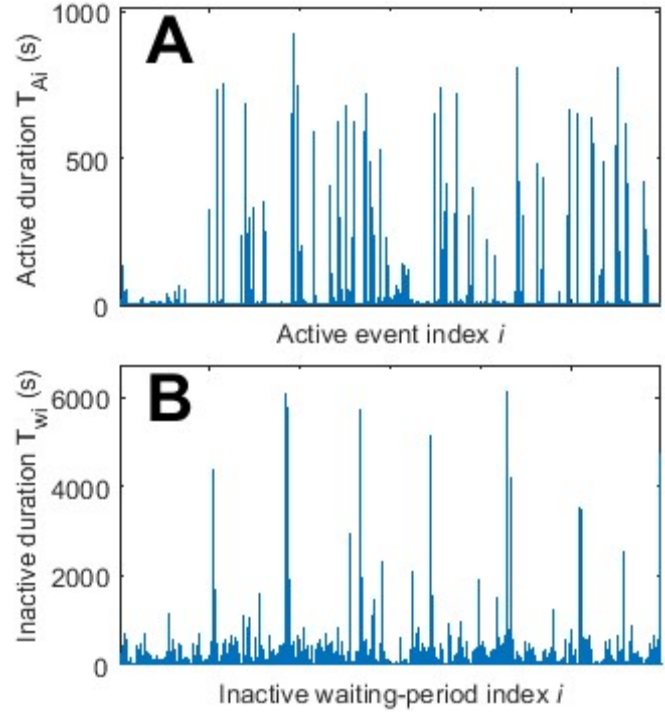

**Figure 2** – Series of events extracted from accelerometer data using a threshold of the mean of non-zero activity counts. Series of (**A**) active events (successive durations above the threshold), and (**B**) inactive waiting-times (successive durations below the threshold). Data above the threshold are active events of duration $T_{A_i} = t_{D_i} - t_{U_i}$, and data below the threshold are inactive waiting-times of duration $T_{W_i} = t_{U_{i+1}} - t_{D_i}$, where $t_{U_i}$ is the beginning (upward threshold crossing), and $t_{D_i}$ is the end (downward threshold crossing), of the $i^{th}$ event.

**Pooled waking and sleeping data**

The BICs for active period CDFs indicate that a composite of exponential and truncated power-law models (Figure 3A) was the most successful model at the individual level (n=54), followed by power-law (n=29), truncated power-law (n=9) and bi-exponential (n=7). Those described by power-law had lower MVPA than those described by the composite model (mean=2% vs. 6%; p<0.001). Visually, the inactive distributions appear to follow the same composite model as the active distributions (Figure 3B); however, after penalizing for model complexity, the simpler truncated power-law form dominates. Most inactive CDFs were described by truncated power-law (n=75); 24 were described by power-law. The truncated power-law model indicates a curtailing of inactive periods at long timescales, and transition into activity. Participant characteristics (age, BMI, sex, psychological distress, smoking status) were similar between these two groups (p>0.052).

Exceedance probabilities quantitatively confirmed that the composite model for the active periods, and the truncated power-law for the inactive periods, were the most successful group-wise models (Table 3).

**Table 3**

**Bayesian model selection and group-averaged parameters for *pooled* waking and sleep data (n=89)[a]**

| Distributions | Active distributions | | Inactive distributions | |
|---|---|---|---|---|
| | Exceedance probability ($\varphi$) | Group-averaged parameters | Exceedance probability ($\varphi$) | Group-averaged parameters |
| Power-law | 0.027 | $\alpha$=2.90 | 0.177 | $\alpha$=2.04 |
| Exponential | 0 | NA | 0 | NA |
| Log-normal | 0 | $\mu$=-2.12<br>$\sigma$=1.88 | 0 | NA |
| Weibull | 0 | $\beta$=0.14<br>$\lambda$=9.48x10$^{-7}$ | 0 | NA |
| Truncated power-law | 0 | $\alpha$=2.02<br>$\lambda$=0.004 | **0.823** | **$\alpha$=1.64**<br>**$\lambda$=2.0x10$^{-4}$** |
| Biexponential | 0 | $\lambda$=0.37<br>$\gamma$=0.03<br>$\delta$=0.86 | 3.3x10$^{-5}$ | $\lambda$=0.007<br>$\gamma$=0.029<br>$\delta$=0.598 |
| Sum of exponential and truncated power-law | **0.973** | **$\alpha$=1.73**<br>**$\lambda$=0.03**<br>**$\gamma$=0.42**<br>**$\delta$=0.48** | 7x10$^{-4}$ | $\alpha$=1.56<br>$\lambda$=0.002<br>$\gamma$=0.001<br>$\delta$=0.099 |

**Note**: Models with group-averaged parameters of "NA" were not potentially plausible models for any cases; plausible models were determined from the log-evidence ratios of at least 0.05.

[a] Ten participants had insufficient sleep data, so were not included in analysis of pooled waking and sleeping data.

Successful group-wise models identified from exceedance probabilities represent the most likely model to describe each individual's CDF within the cohort, given the set of competing models. That is, a composite group-wise model reflects a winning composite model at the individual level, not a composite of different individual-level simple models. Multiple hierarchical regressions indicated that MVPA was a significant predictor for α, λ and μ of the composite model describing active periods; SB was also a significant predictor of α. For the truncated power-law model describing inactive periods, SB was a significant predictor for α and λ, with MVPA and smoking status also predicting α, and age also predicting λ. These results are shown in Supplementary Table S2.

CDFs generated from 60-sec epoch data (Supplementary Figure S1) were described by the simpler truncated power-law model (φ =1.0), and lacked the two distinct regimes evident in 1-sec data (compare Figure 3 and Supplementary Figure S1).

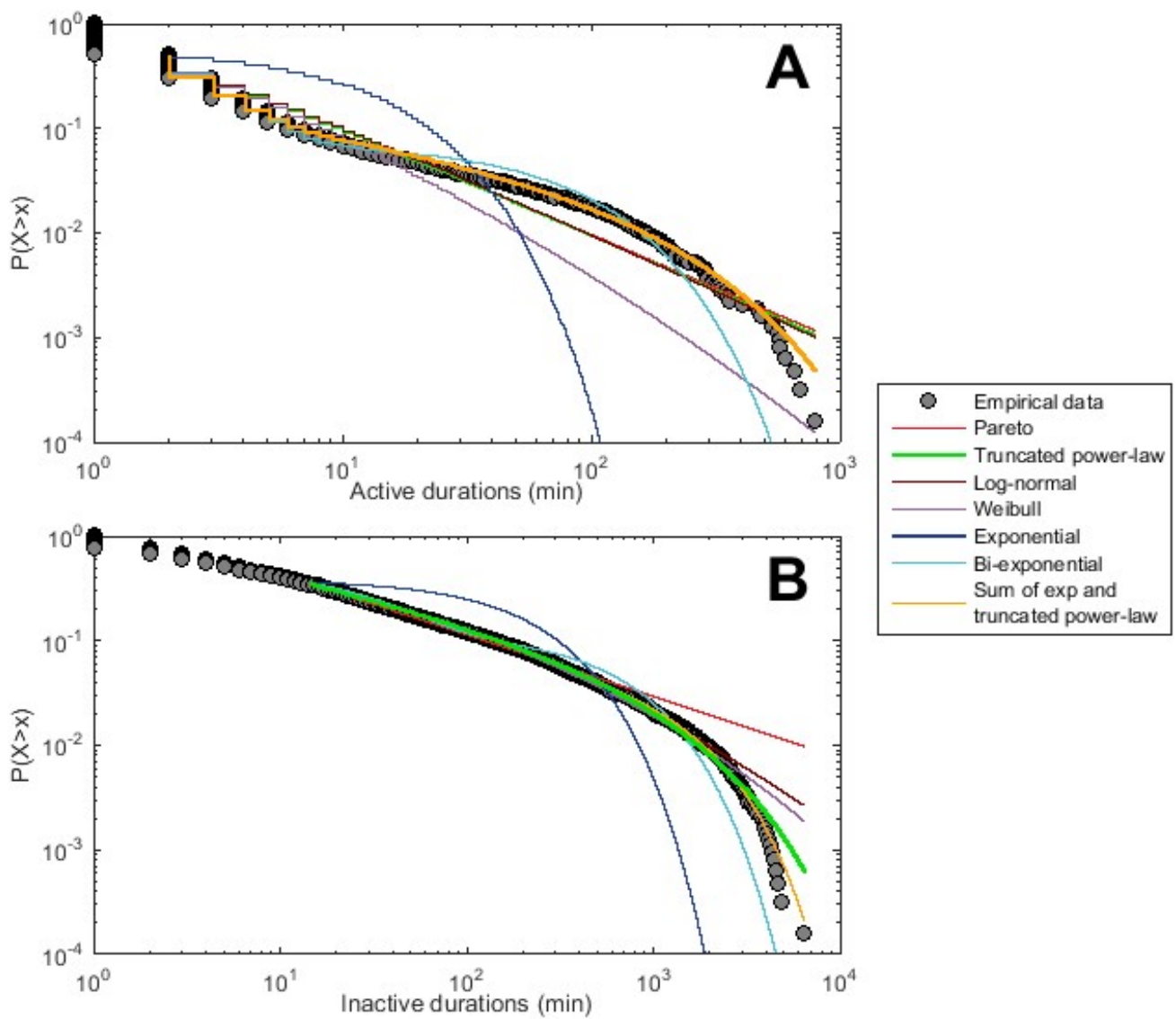

**Figure 3** – Cumulative distribution functions (CDFs) and model fits for (**A**) active durations, and (**B**) inactive durations. The most successful group models (denoted by thick lines) were the sum of exponential and truncated power-law for active CDFs (parameters: α=1.73, λ=0.03, γ=0.42, δ=0.48), and truncated power-law for inactive CDFs (parameters: α=1.64, λ=0.0002).

## Waking vs. sleeping data

Similar characteristic behavior was found for participants' waking patterns (Figure 4): A sum of exponential and truncated power-laws described active distributions ($\varphi=0.961$; Figure 4A), and the truncated power-law described inactive distributions ($\varphi=0.988$; Figure 4B). Sleep exhibited an absence of heavy-tailed behavior for active distributions, following the simple exponential form ($\varphi=1.0$; Figure 4A). Inactive distributions during sleep were more complex, being described by the sum of exponential and truncated power-laws ($\varphi=1.0$; Figure 4B). Model statistics are shown in Supplementary Table S3.

During waking hours, the mean proportion of daily time spent in MVPA and SB was 4.5% (SD=4.5%) and 65% (SD=11%), respectively. Other summary statistics of waking behaviour have been presented elsewhere[47].

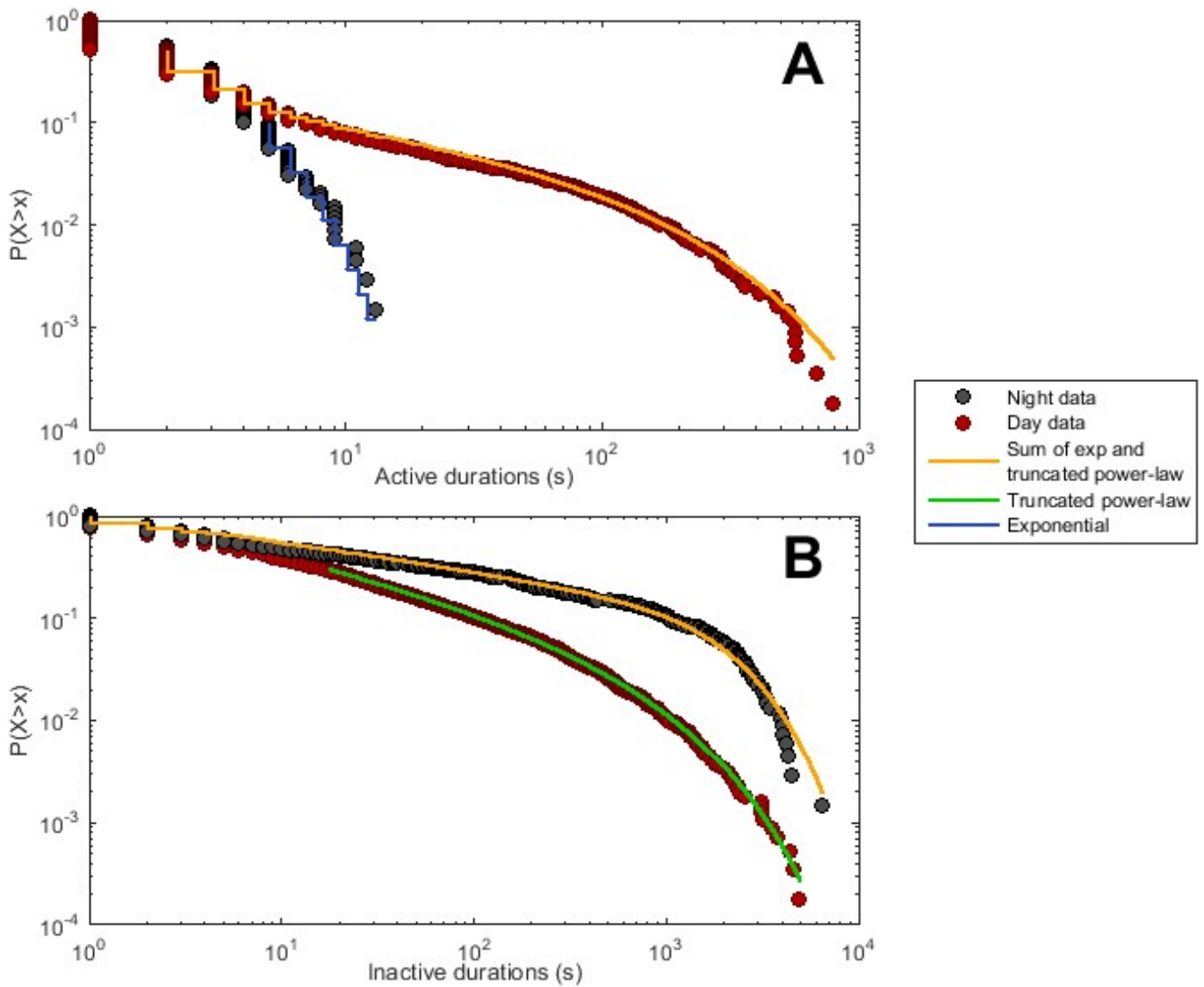

**Figure 4** – Cumulative distribution functions (CDFs) for (**A**) active durations, and (**B**) inactive durations, for self-reported time awake (red circles), and time in bed at night (black circles). Only the successful group models are shown.

**Subgroup comparisons**

Prior research and clinical observations suggest that different clinical disorders may be characterized by distinct differences in activity patterns[48]. We compared activity patterns between participants with psychotic and bipolar disorders. Participants in these groups had similar age (mean=45.1 vs. 45.9; p=0.75), BMI (mean=31.8 vs. 31.1; p=0.72), distress (mean=11.5 vs. 13.4; p=0.23), SB (mean=65% vs. 62%; p=0.47), and sex (female=31% vs. 50%; p=0.26), but participants with psychotic disorders had higher MVPA than those with bipolar disorder (mean=2.8% vs. 8.4%; p=0.003). A composite of exponential and power-law models described psychoses, whereas a simpler power-law model described bipolar disorder (Figure 5). We further confirmed that these two groups were generated from different distributions using the log Bayes factor (LogBF=13.2, F-statistic=7.5, corresponding to strong evidence).

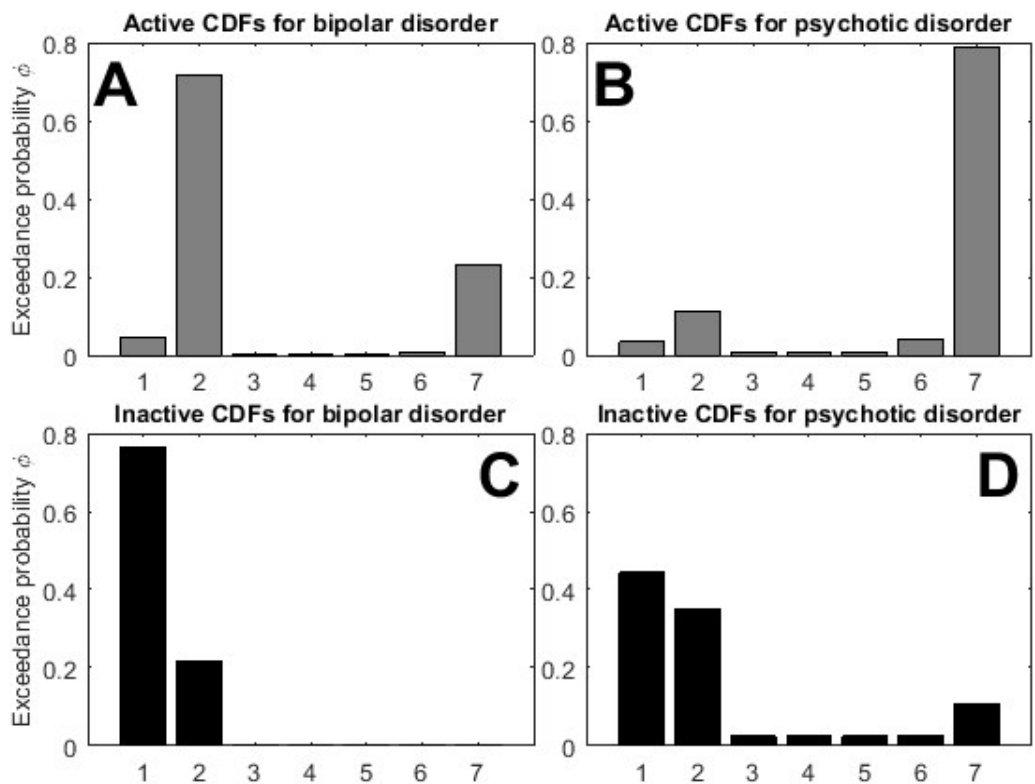

**Figure 5** – Exceedance probabilities ($\varphi$) of the seven competing models fit to cumulative distribution functions (CDFs): **1**=truncated power-law; **2**=power-law; **3**=log-normal; **4**=Weibull; **5**=exponential; **6**=biexponential; **7**=sum of exponential and truncated power-law. Upper panels (**A, B**) represent exceedance probabilities for active distributions; lower panels (**C, D**) represent exceedance probabilities for inactive distributions. Left panels (**A, C**) show results of participants with bipolar disorders (n=14), right panels (**B, D**) show results of participants with psychotic disorders (n=13). Active distributions were best described by a power-law distribution for participants with bipolar disorders (exceedance probability $\varphi$=0.77; parameter: $\alpha$=3.19), and a sum of exponential and truncated power-law for participants with psychotic disorders (exceedance probability $\varphi$=0.79; parameters: $\alpha$=1.85, $\lambda$=0.01, $\gamma$=0.55, $\delta$=0.47). Inactive distributions were similar for both groups, being described by truncated power-laws, with parameters: $\alpha$=1.62, $\lambda$=0.0002, and $\alpha$=1.69, $\lambda$=0.0002, for participants with bipolar and psychotic disorder, respectively.

Physical activity and metabolic health have been increasingly recognized as crucial in people with psychoses[49]. We thus compared the group-averaged model parameters between obese (BMI>32; n=16) and non-obese (BMS<32; n=12) participants with psychotic disorders, using a Mann Whitney-U test. These participants had similar age (mean=44 vs. 38; p=0.07), distress (mean=13 vs. 16; p=0.63), SB (mean=65% vs. 65%; p=0.91), MVPA (mean=3% vs. 5%; p=0.21) and sex (female=55% vs. 40%; p=0.22). The truncation parameter was significantly higher for obese ($\lambda$=0.02) than non-obese participants ($\lambda$=0.003; *p*=0.03).

**Threshold variation**

To verify that our results are insensitive to the exact choice of the threshold, we varied the threshold over a large range, from 0.5 to 1.5 times the mean of non-zero counts (Figure 6). The inactive period CDFs, and corresponding parameter estimates, are robust over a wide range of threshold values. On average, the scaling parameter $\alpha$ changed by less than 10% over the range of threshold values. More variation was observed for the active period CDF and model parameters: on average, the scaling parameter $\alpha$ changed by less than 50% over the range of threshold values.

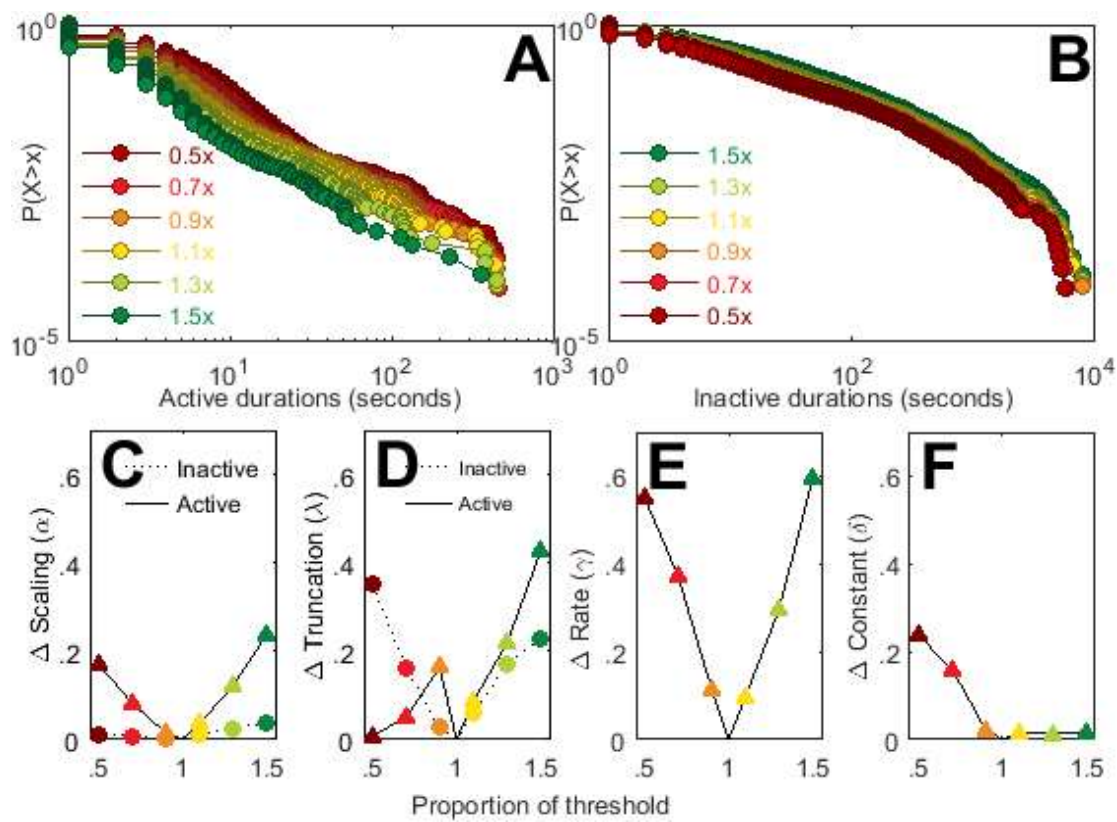

**Figure 6** – Robustness of cumulative distribution functions (CDFs) to variation in the activity threshold. The chosen threshold was defined as the mean of non-zero activity counts, and varied by +/- 50%. Data stability is shown for (**A**) active CDFs, and (**B**) inactive CDFs. Stability of the successful model parameters shown for (**C,D,E,F**) active distributions (parameters $\alpha$, $\lambda$, $\gamma$, and $\delta$ of sum of exponential and truncated power-law), and (**C,D**) inactive distributions (parameters $\alpha$ and $\lambda$ of truncated power-law); y-axes represent the parameter's proportional change from the chosen threshold.

**DISCUSSION**

Monitoring free-living physical activity patterns has important applications in psychiatry, for which mobile phones offer a powerful "real world" medium for data acquisition. However, validation of robust analytical techniques for the characterization of movement patterns is needed prior to broader implementation. Using a novel analytical method to analyze accelerometer data from adults with mental illness, we demonstrated that the stochastic models that best describe behavior are more complex than previously thought. This approach can be applied to accelerometer data acquired from mobile phones, and may have broader applications in monitoring severity and progression of other psychiatric and neurological disorders, such as dementia, Alzheimer's[50] and Parkinson's disease[51].

Our main finding was that free-living activity in our heterogeneous clinical sample was described by a previously unrecognized composite of two distinct regimes of behavior. Short duration (~10 s) exponential behavior may be caused by externally-triggered reactions to environmental stimuli, whereas longer duration (~1000 s) heavy-tailed activity may be caused by internally-driven cognitive processes responsible for task priority queuing[32]. Our data suggests that both movement and the cessation of movement have characteristic time scales: Inactive distributions spanned four orders of magnitude (1-se to ~2.5 hours), and active distributions spanned three orders of magnitude (1-sec to ~17 minutes). Here, truncation of active distributions may be caused by the energetic extent of the system (e.g. fatigue), or the completion of tasks, resulting in cessation of meaningful activity. During periods of rest, the need to respond to internal and external influences (e.g. hunger, adjustments while seated, the development of a task priority) prevail at long timescales, and transition into activity. These findings help constrain the type of generative dynamics occurring in the cortical areas associated with motor preparation of self-initiated movement, such as the anterior mid-cingulate cortex and the supplementary motor area[52]. Whereas uncorrelated, random events yield exponential CDF's, the presence of long-tailed statistics in time series data speaks to more complex underlying processes, such as temporal memory, metabolic constraints and strong feedback[35]. These findings thus help constrain the type of underlying dynamics occurring in the cortical areas associated with motor preparation of self-initiated movement such as the anterior mid-cingulate cortex and the supplementary motor area[52]. In particular, while the neural substrates for these areas are well known, the dynamics that arise from their interactions – and that support movement onset and maintenance – are not well known. The presence of long-tailed statistics and composite statistical models suggests that the underlying dynamics occur in the presence of nonlinear instabilities and, possibly, multi-attractor systems[35,53,54]: This is an intriguing area that warrants further study.

Previous research has found that power-law and stretched exponential (Weibull) models best describe the cumulative distributions of inactive and active periods, respectively[36]. Differences between previous research and our findings may arise from the limited number of previous models studied, the absence of a formal inversion scheme, and the different time scales used (data averaged over 30-second[55], or 60-second

epochs[29,36-38]) which appears to be insensitive to this composite behavior. The position of accelerometer wear also influences the resultant activity pattern[56]: we used hip-worn monitors which measure "whole body" movement (and may more closely mimic the position of a mobile phone), whereas previous studies have used wrist-worn devices.

We also found that heavy-tailed behavior was not evident during sleep. The observed dominance of exponential behavior during sleep may support the preceding hypothesis, because higher order task queuing is removed during sleep, superseded by environmental responses (e.g. uncomfortable sleep position), or lower order cognitive processes (e.g. dreams). The distinction between 'sleep' and 'awake' data, however, was made from self-reported time to bed and time out of bed, which does not consider waking periods in bed (e.g. sleep latency, awakenings), or napping during the day. The Bayesian model selection routine is robust to outliers, so participants with particularly restless sleeps, or extensive daytime napping, are not likely to have altered the group-wise model selection.

The utility of these findings is exemplified by our application of a comprehensive set of competing models to the data, and the quantification of group-based likelihood of the models using a Bayesian approach. Previous studies have applied a smaller number of model hypotheses to activity data, which can lead to misinformed conclusions[57], and have not formally appraised group-wise model success. We used a threshold defined by the mean of activity counts to define the time-series of activity or inactivity, consistent with previous literature[29,36,37]. This approach may be advantageous because the threshold is driven by the data, and may therefore be more sensitive to differences in activity patterns. However, other widely accepted accelerometer statistics use externally-validated thresholds that correspond with categories of energy expenditure. Whilst useful for determining the time spent at different intensities of activity, these thresholds are dependent upon the method of validation (e.g. lab-based vs. free-living) and individual characteristics (e.g. age, sex, BMI)[58], which can produce large discrepancies in outcomes[59]. Identification of potentially abnormal activity patterns using our presented analyses would require comparison with a control group matched on the time spent in physical activity. Future research could investigate threshold-free analyses.

We additionally performed a proof-of-principle contrast between participants with different psychiatric diagnoses. Interestingly, the short-duration exponential behavior evident in the active distributions of participants with a psychotic disorder was absent in those with bipolar disorder. Instead, a simple power-law model best governed the active distributions in bipolar patients, potentially representing a reduction in reactional behavior(Figure 5). This finding is broadly consistent with the hypothesis of neuronal dysregulation of dopamine in bipolar disorder, causing over-excitation of the processes responsible for volition[60]. Previous research has found that bipolar patients in a manic episode exhibited a greater degree of exploratory behavior when faced with a novel environment[7,61], which may be reflected in the present findings, and may have clinical implications for the monitoring of psychopathological symptoms. Interestingly, participants with psychotic disorders had a higher level of moderate-to-vigorous activity than bipolar patients. This may seem contradictory with the preceding hypothesis, however, the composite model describing activity patterns of participants with psychoses curtails active periods at long durations indicating a 'ceiling' effect, whereas the power-law describing those with bipolar disorder does not have a characteristic time-scale. Future research should compare participants with different psychiatric diagnoses matched by physical activity level to verify if the difference in activity patterns between diagnostic groups are independent of activity level.

Adults with mental illness have poor physical health compared with the general population[62]. Changes in activity patterns may reflect a higher risk of developing physical complications. We examined a limited number of parameters pertaining to physical health, and found that the model statistics for obese people with psychotic disorders were different from those who were not obese. A larger truncation parameter was found for the active distributions of those who were obese, indicating a reduced likelihood of remaining active for longer periods of time. This could be due to a higher propensity to fatigue in obese participants with a psychotic disorder, possibly driven by 'negative' symptomatology. The application of this analytical method is formative, however, and a larger sample would be required to investigate the possible influences of health and demographic characteristics on model parameters.

Our data were recorded during free-living conditions, sampled from a diverse sample of adults across a range of mental illnesses. While this approach has strong ecological validity, it also presents a number of limitations. In particular, there were only 14 participants that had a single psychiatric diagnosis without comorbidities, which limits investigation into the specific influence of diagnosis, medications, and demographic characteristics. Sixteen participants did not meet the MINI-Plus criteria for a current mental illness (as indicated in Table 1, note c); however, these participants may have had symptoms of poor mental health that did not reach clinical levels. The sample heterogeneity may also have advantages, in that the application of this analytical method to a broad sample is more reflective of real world conditions.

Future work could focus on applying these analytical techniques to larger samples of participants with a specific psychiatric diagnosis, and correlating model parameters with dimensional symptomatic measures (e.g. positive and negative symptoms of people with schizophrenia) rather than contrasts between diagnostic categories. A previous study reported that avolition, but not qualitative motor disturbance, was associated with reduced activity levels in adults with schizophrenia[21]; analyses presented here quantify the complete activity distribution, and may have utility for detecting motor disturbances[38]. Our analyses could also be used to quantify sleep disturbances. Indictors of disturbed motor activity or sleep could be used as early warning signs of relapse and to promote early intervention, or to assess treatment progression or efficacy. We analysed pooled waking and sleeping data, and waking and sleeping data separately; this technique could also be applied to shorter time periods (e.g. ~1-hr) similar to other research[63]. Further, the parameters of the active and inactive distributions for each participant may correlate with over- or under-activity of different brain regions, similar to previous research[64]. Future work should compare activity patterns with a healthy comparison group to detect potentially abnormal movement patterns associated with particular illnesses.

In summary, we presented a novel approach that may have applications in the monitoring of pathophysiological symptoms related to movement abnormalities in adults with mental illness. This methodology holds promise for broader application, particularly with the roll-out of mobile phone-based psychiatry research.

## ACKNOWLEDGEMENTS

The authors would like to thank the Royal Brisbane and Women's Hospital, The Prince Charles Hospital, Communify QLD, Footprints Inc., Mental Illness Fellowship QLD, Brooke Red, and Reclink Australia for their support and assistance with recruitment. This work was not funded from any specific grant or sponsor.

## AUTHOR CONTRIBUTIONS

JC collected all data, performed analyses, and prepared the manuscript text and figures. JR provided advice and direct assistance with analyses. VN provided advice on analyses. MB advised on project design, and directed analyses. All authors reviewed and edited the manuscript.

## COMPETING FINANCIAL INTERESTS

The authors Justin Chapman, Dr James Roberts, Dr Vinh Nguyen and Professor Michael Breakspear declare that there are no biomedical or financial interests, and there are no potential conflicts of interest. This research received no specific grant from any funding agency in the public, commercial, or not-for-profit sectors.